\documentclass{article}
\pdfoutput=1
\usepackage[utf8]{inputenc} 
\usepackage[T1]{fontenc}    
\usepackage{hyperref}       
\usepackage{url}            
\usepackage{booktabs}       
\usepackage{amsfonts}       
\usepackage{nicefrac}       
\usepackage{microtype}      
\usepackage{lipsum}			
\usepackage{graphicx}
\graphicspath{ {../output/} }
\usepackage[numbers]{natbib}
\usepackage{doi}
\usepackage{caption}
\usepackage{subcaption}
\usepackage{authblk}
\usepackage{float}
\usepackage{etoolbox}
\usepackage[nobottomtitles*]{titlesec}

\providecommand{\keywords}[1]
{
  \small
  \emph{\textit{Keywords---}} #1
}

\title{Early Warning Software for Emergency Department Crowding}

\def\correspondingauthor{\footnote{
	Corrersponding author.

	\emph{Email address:} jalmari.tuominen@tuni.fi
}}

\author[1]{Jalmari Tuominen\correspondingauthor{}}
\author[4]{Teemu Koivistoinen}
\author[2]{Juho Kanniainen}
\author[1,3]{Niku Oksala}
\author[1,4]{Ari Palomäki}
\author[1]{Antti Roine}

\affil[1]{\footnotesize Faculty of Medicine and Health Technology, Tampere University}
\affil[2]{\footnotesize Faculty of Information Technology and Communication Sciences, Tampere University}
\affil[3]{\footnotesize Vascular Centre, Tampere University Hospital, Tampere, Finland}
\affil[4]{\footnotesize Kanta-Häme Central Hospital, Hämeenlinna, Finland}

\date{}

\hypersetup{
pdftitle={Early Warning Software for Emergency Department Overcrowding},
pdfsubject={eess.SY},
pdfauthor={Jalmari Tuominen, Teemu Koivistoinen, Juho Kanniainen, Niku Oksala, Ari Palomäki, Antti Roine},
pdfkeywords={Emergency department, Crowding, Overcrowding, Forecasting, Prospective, Software, ETS models},
}

\begin{document}
\maketitle

\AtBeginEnvironment{tabular}{\footnotesize}

\begin{abstract}
Emergency department (ED) crowding is a well-recognized threat to patient safety and it has been repeatedly associated with increased mortality. Accurate forecasts of future service demand could lead to better resource management and has the potential to improve treatment outcomes. This logic has motivated an increasing number of research articles but there has been little to no effort to move these findings from theory to practice. In this article, we present first results of a prospective crowding early warning software, that was integrated to hospital databases to create real-time predictions every hour over the course of 5 months in a Nordic combined ED using Holt-Winters' seasonal methods. We showed that the software could predict next hour crowding with a nominal AUC of 0.98 and 24 hour crowding with an AUC of 0.79 using simple statistical models. Moreover, we suggest that afternoon crowding can be predicted at 1 p.m. with an AUC of 0.84.

\keywords{Emergency department, Crowding, Overcrowding, Forecasting, Prospective, Software, ETS models}
\end{abstract}

\section{Introduction}\label{introduction}
Emergency department (ED) crowding is a well-recognized threat to patient safety and it has been repeatedly associated with increased mortality \cite{Richardson2006, Berg2019, Jo2014, Guttmann2011, Sprivulis2006}. Crowding is both a chronic and an international issue but despite a multitude of studies and media coverage, the problem seems to be getting worse \cite{Janke2022}. According to a conceptual model, causes of ED crowding can be divided into three high-level components: input, throughput and output \cite{Asplin2003}. In a recent review article by Morley et al, three respective phenomena were highlighted as underlying causes for crowding: 1) increased number of patients with more urgent and complex care needs, 2) nursing staff shortages and 3) access block (i.e. difficulty to move patients from the ED to follow-up care after initial assessment and immediate treatment) \cite{Morley2018}. From the point of view of an individual ED administrator, it is difficult to influence these factors, since majority of the required interventions are locked behind slow political processes scattered throughout the health care system.

For these reasons, there is a continued interest aiming to enhance the use of the limited resources that are readily available. One notable manifestation of this effort is emergency department forecasting, which has established itself as a small but persistent research niche \cite{Gul2020}. The rationale of the forecasting work is simple: 1) forecast future service demand, 2) enable proactive administrative decisions, 3) ensure sufficient resources and 4) improve treatment outcomes.

Despite many recent advancements in forecasting methodology \cite{Caldas2022, Harrou2020, Sharafat2021, Fan2022}, several gaps in the literature remain. First, there is little to no knowledge about the performance of the forecasting models in predicting future crowding in binary terms. This is because the models are predominantly assessed using continuous error metrics such as mean absolute percentage error (MAPE) \cite{Gul2020}. These metrics are useful when comparing forecasting models between each another, i.e. analyzing whether model $A$ is better than model $B$. However, compared to classification metrics, continuous error metrics can be difficult to use to assess if the shift-supervising physician can base decisions on the model. Therefore, the performance of forecasting models should be assessed using discrete metrics that are easily interpretable to administrative stakeholders of the ED's. From practitioners' point of view, the day-to-day ED administration is interested whether model $A$ or $B$ can predict if next hour, next day or the next seven days will be crowded rather than drilling into the continuous error metrics to analyse whether model $A$ is better than model $B$ under given circumstances.

ED crowding is studied as a discrete phenomenon also from an academic standpoint. Particularly, many of the studies that have documented the association between mortality and crowding have done so by comparing the most crowded quartile between less crowed ones \cite{Richardson2006, Berg2019, Jo2014}. Although the underlying association between increased occupancy and mortality is likely a continuous one, this kind of categorization is beneficial from practical standpoint, since decision-making is known to benefit from actionable, simple output \cite{Elstein2002}.

In Tampere University Hospital, following the rationale above, the ED is considered crowded when a certain occupancy threshold is exceeded. This is coupled with a catchment-area-wide protocol that aims to resolve the observed crowding. The protocol mandates the shift-supervising physician to call-in additional staff and obligates follow-up care facilities to accept patients even if their nominal capacity has been exceeded. The obvious problem with this approach is the delay between actions and outcomes, which leads to prolonged crowded state, increased length of stay and decreased quality of care. Our ultimate goal is to move these administrative manoeuvres from reactive space into proactive space by offering accurate forecasts about the future status of the ED.

Returning to the gaps in current literature, virtually all the previous studies have reported results based on historical simulations. Retrospective studies are obviously useful in identifying potential forecasting models and identifying operational bottlenecks. However, the impact of retrospective studies alone will remain limited unless their findings are confirmed prospectively in real-life setting. Second, ED forecasting models have been under examined in the Nordic countries with few notable exceptions \cite{Ekstrom2015}, \cite{Asheim2019}. Third, increasingly complex solutions are reported retrospectively \cite{Caldas2022, Harrou2020, Sharafat2021} but since they are only evaluated in continuous terms, it is difficult to assess whether these models have practical utility. Meanwhile, even the most rudimentary statistical models remain unassessed in discrete terms.

In this study, we were set out to fill the gaps identified above. We developed a software solution that was integrated to Tampere University hospital databases to make hourly predictions of ED arrivals and occupancy 24 hours ahead using established statistical models. In this study, we report the performance of the system in predicting discrete future crowding along with its reliability.

\section{Materials and Methods}\label{materials_and_methods}
\subsection{Study setting}

Tampere University Hospital is an academic hospital located in Tampere, Finland. It serves a population of 535,000 in the Pirkanmaa Hospital District and, as a tertiary hospital, an additional population of 365,700, providing level 1 trauma centre capabilities. The hospital ED, \emph{Acuta}, is a combined ED with a nominal capacity of 106 patients, with 65 beds and 41 seats for walk-in patients. Approximately 90,000 patients are treated annually making \emph{Acuta} one of the largest EDs in Scandinavia.

For the purposes of this study, our research group developed a forecasting software that had following requirements: 1) it had to be able to forecast both future arrivals and occupancy, 2) it had to operate with hourly data, making predictions 24 hours ahead, 3) the predictions had to be stored in a database for later accuracy evaluation and 4) it had to have a rudimentary user interface for debugging purposes. The software was deployed to an Azure cloud computing service (Microsoft Corporation, USA) on January 15, 2022 and predictions were made until May 26, 2022 constituting a total of 3145 hours. To ensure safety of sensitive patient information the virtual machine was isolated from hospital databases and only the high-level statistics required to make predictions were provided to a dedicated database.

\subsection{Models}
The software included three forecasting models, Holt-Winter's additive method (AHWM), Holt-Winters' multiplicative method (MHWM) and Holt-Winters' damped method (DHWM) \cite{Winters1960}. These models were selected due to their established status, capability to process seasonal data and efficiency in terms of computing power. The models, later referred collectively to as ETS models (E stands for error term, T for trend component, and S for seasonal component) were trained with all available historical data and implemented with Statsmodels Python module \cite{statsmodels}. ETS model parameters were determined using maximum likelihood estimation.

\subsection{Performance metrics}

 The performance of these algorithms are reported in three distinct phases: (1) aggregated continuous performance, (2) performance as a function of a forecast horizon and, (3) performance as a function of a forecast origin. Only out of sample performance results are presented in this study. In the first phase, the continuous performance of the models is reported using mean absolute error (MAE) and root mean squared error (RMSE). The errors are averaged over all the forecast horizons.

In the phases two and three, we evaluated the performance of the models as binary predictors of future crowding. In these phases, the ED was considered overcrowded if the absolute occupancy reached the highest occupancy quartile. Area under the receiver operating characteristics curve (AUROC/AUC) were used as the main error metric in both phases and detailed unadjusted binary metrics, such as F1, are provided in the Appendix. F1 is an unweighted harmonic mean of precision and sensitivity (recall), which is often recommended as a performance measure under class imbalance \cite{forman2010apples}.

\emph{Performance as a function of forecast horizon (PFFH)} is documented in phase two. This allows the reader to assess how well the model is able to predict whether the ED will be crowded exactly \(t\) steps ahead. For example, $t+2$ PFFH for predictions made at 1 p.m. tells whether the ED will be crowded between 3 p.m. and 4 p.m. Each horizon from 1 to 24 hours are independently assessed. The issue with PFFH metric is that the difficulty of forecasting future crowding is intuitively affected by the forecast origin. This is because hourly occupancy follows a clear seasonal pattern and hence most of the crowding events are observed in the afternoon. It is thus intuitively relatively easy for the model to predict that \(t+1\) will be overcrowded if the forecasts are performed e.g. at 6 p.m. when the ED is often already crowded.

\emph{Performance as a function of forecast origin (PFFO)} in phase three resolve this issue. In this phase we assess whether the models are able to tell whether the next 24 hours will be overcrowded as the forecast origin moves from 0 a.m.  to 23 p.m. We hypothesize that the accuracy is lowest at midnight and gradually increases throughout the day when the models is given new information about the status of the ED. The approach aims to simulate the way the forecasts would be used in practice if offered to ED administration. In this phase, the future is considered crowded if the occupancy reaches the highest quartile for one or more hours within the 24 hour forecast window.

\section{Results}\label{results}
\subsection{Descriptive statistics}
\begin{figure}[ht]
     \centering
     \begin{subfigure}[b]{0.33\textwidth}
         \centering
         \includegraphics[width=\textwidth]{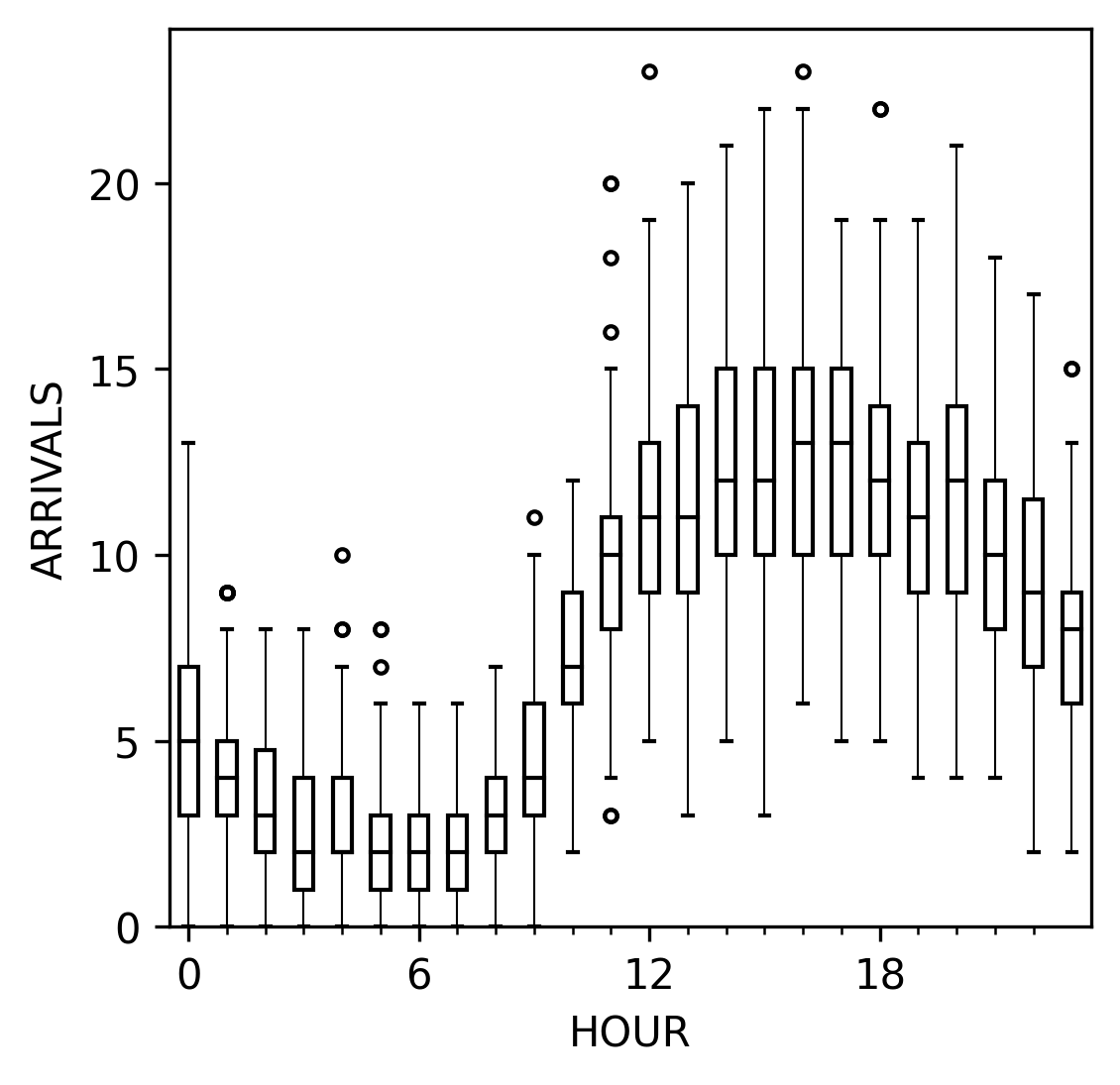}
         \caption{Arrivals}
         \label{fig:season_arrivals}
     \end{subfigure}
     \begin{subfigure}[b]{0.33\textwidth}
         \centering
         \includegraphics[width=\textwidth]{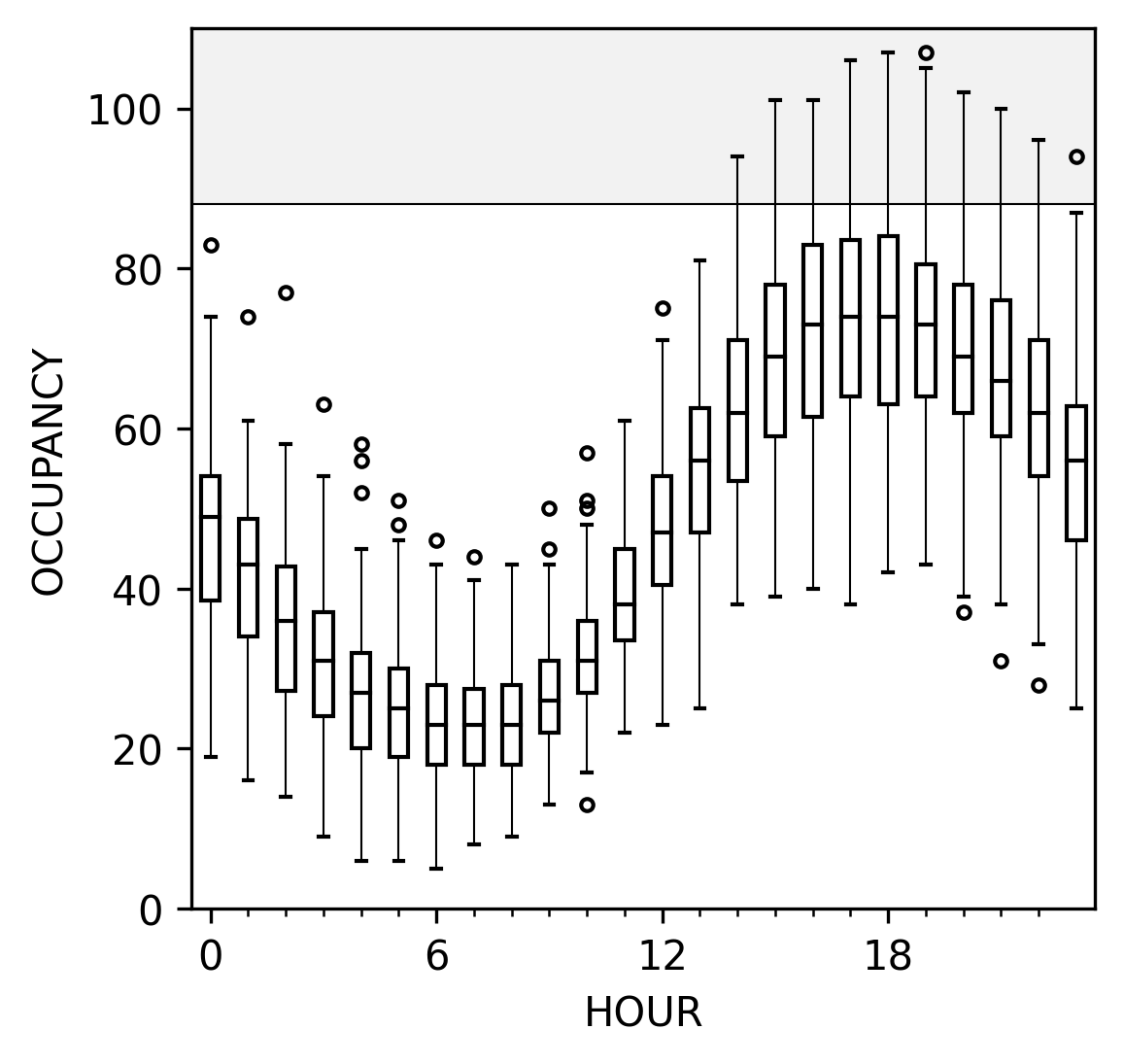}
         \caption{Occupancy}
         \label{fig:season_occupancy}
     \end{subfigure}
     \hfill
     \caption{Hourly target variable seasonalities. The grey area shows the overcrowded state.}
     \label{fig:target_seasonalities}
\end{figure}

\begin{figure}[ht]
    \centering
    \includegraphics[width=0.49\textwidth]{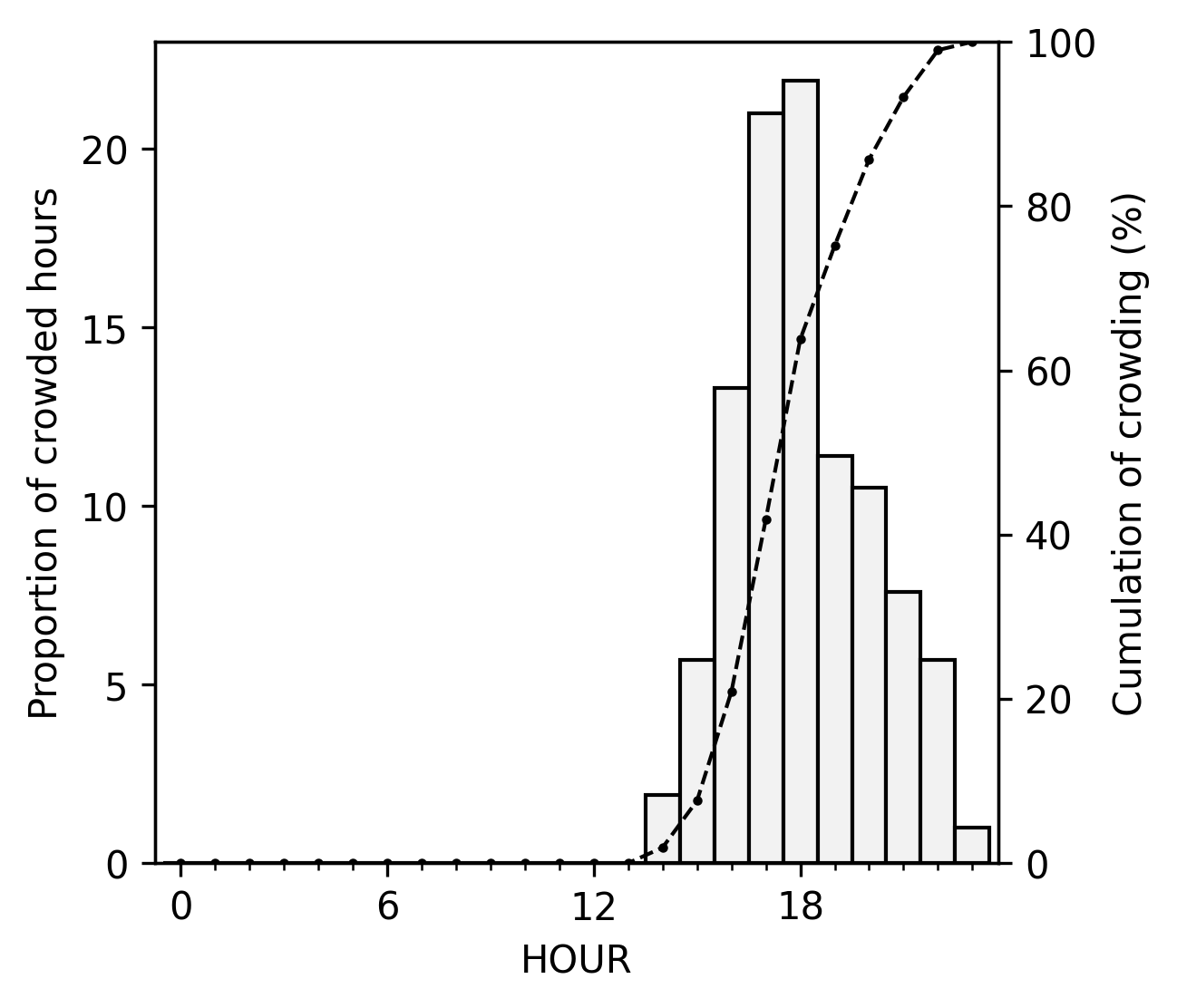}
    \caption{Temporal distribution of crowding events as a function of hour of the day. Dotted line shows the cumulative sum.}
    \label{fig:crowding_dist_hourly}
\end{figure}

Hourly target variables followed a clear seasonality as shown in Figure \ref{fig:target_seasonalities}. On average there were only 2 hourly arrivals between 5-7 a.m. with gradual increase over the course of the morning and throughout the day. Arrivals peaked at 4 p.m. with median of 13 patients and maximum of 23. Occupancy followed a similar sinusoidal shape with slight delay compared to arrivals. The occupancy was lowest between 6-8 a.m. with median of 23 patients and peaked between 5-6 pm. with median of 74 and maximum of 107 patients. Temporal distribution of crowding events, shown in Figure \ref{fig:crowding_dist_hourly}, was a direct result of the seasonality described above. On hourly resolution, all the crowding events were observed after 2 pm and most of them at 6 pm (22 \%). The threshold for the highest occupancy quartile was observed at 88.

The ED visit statistics were delivered reliably to the dedicated database and there were no missing data as regards to target variables. However, this was not the case with the predictions. There was a downtime of two weeks from February 14th to 27th and three days from April 12th to 14th due to a credentials issue. This resulted in total missing data of 397 hours (12\%) for all of the models. In addition, due to an unresolved issue with memory management, the  software experienced sporadic downtime resulting in missing data. This contributed to an additional modelwise data loss ranging between 82-108 (\~3 \%). Temporal distribution of missing data is shown in the Appendix \ref{missing_data}

\subsection{Continuous performance}
\begin{figure}[ht]
    \centering
    \begin{subfigure}[b]{0.49\textwidth}
        \includegraphics[width=\textwidth]{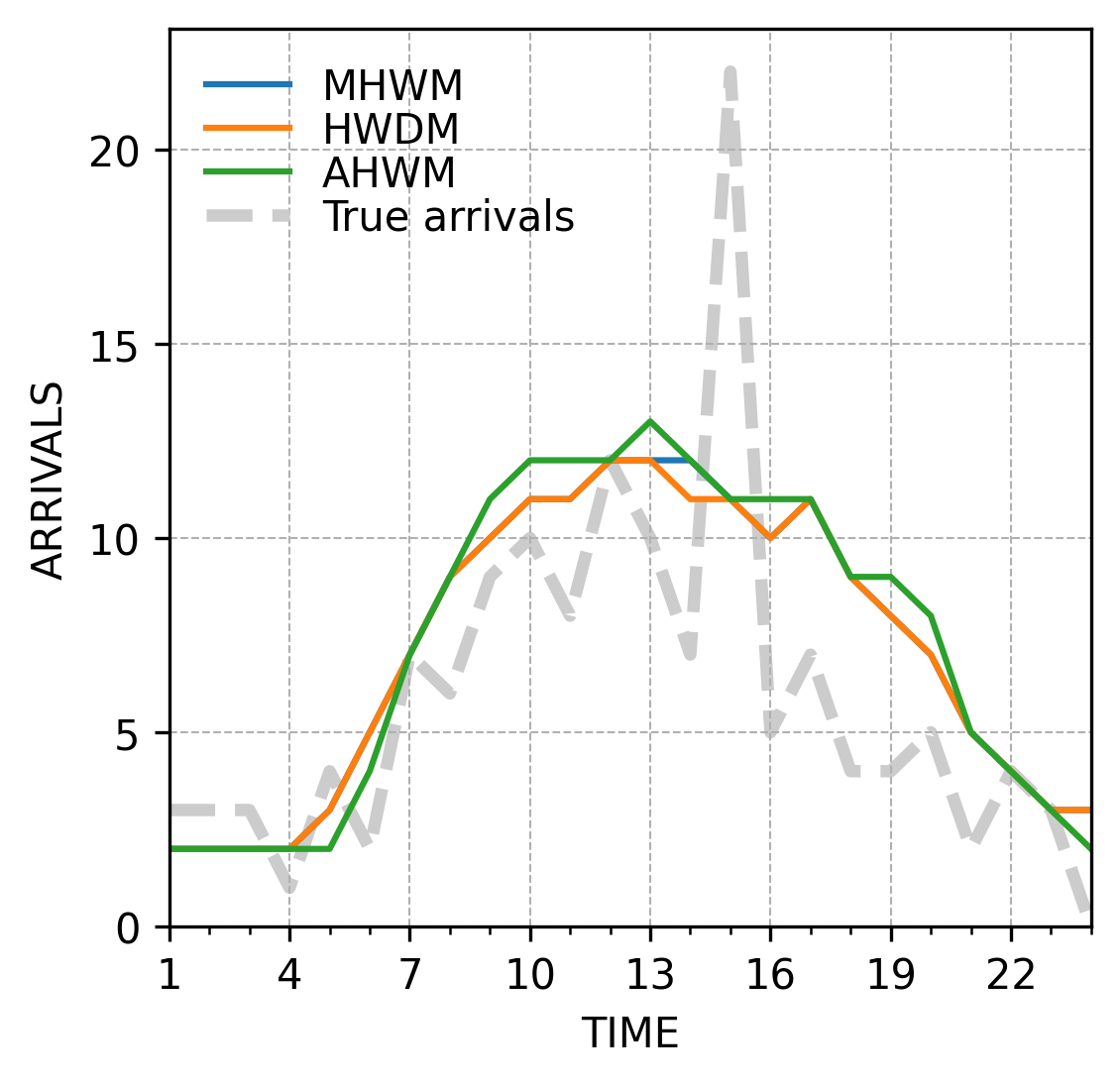}
        \caption{Arrivals}
        \label{fig:example_arrivals}
    \end{subfigure}
    \hfill
    \begin{subfigure}[b]{0.49\textwidth}
        \includegraphics[width=\textwidth]{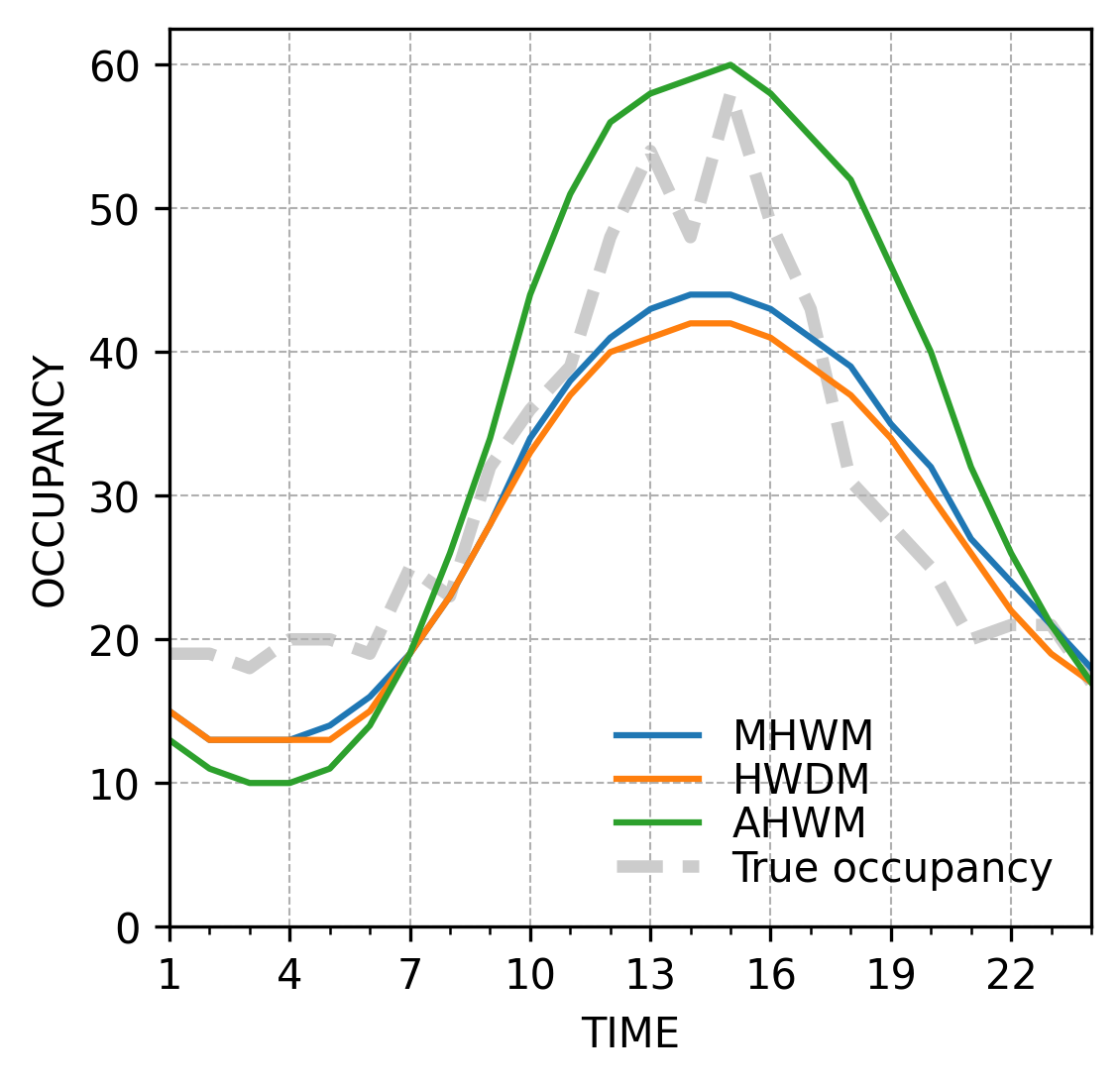}
        \caption{Occupancy}
        \label{fig:example_occupancy}
    \end{subfigure}
    \caption{Example predictions on 2022, February 13th at 4 a.m. Note the almost identical performance when forecasting arrivals, which is also reflected in the overall accuracy results. In case of occupancy, AHWM is closest to the ground truth whereas other ETS models a clear negative bias.}
    \label{fig:example_predictions}
\end{figure}

Continuous accuracy metrics are provided in Table \ref{table:cont_hourly_error}. AHWM was the most accurate model in terms of continuous occupancy with MAE of 10.22 and RMSE of 176.59. The RMSEs of HWDM and MHWM were 35\% and 24\% higher than that of AHWM likely because of their negative bias as demonstrated in Figure \ref{fig:example_predictions}, which unquestionable favours AHWM. In contrast, when it comes to the prediction of arrivals, the models were almost of equal performance, HWDM being slightly the most accurate with MAE of 2.10 and RMSE of 7.58. An example of the predictions is provided in Figure \ref{fig:example_predictions}.

\begin{table}[h]
\centering
\caption{Continuous aggregated error metrics reported with Mean Absolute Error (MAE) and Root Mean Squared Error (RMSE)}
\label{table:cont_hourly_error}
\begin{tabular}{rrcc}
\toprule
         &      &   MAE &   RMSE \\
Target & Model &       &        \\
\midrule
Occupancy & HWDM & 11.69 & 273.47 \\
         & MHWM & 10.87 & 233.38 \\
         & AHWM & 10.22 & 176.59 \\
Arrivals & AHWM &  2.13 &   7.64 \\
         & MHWM &  2.11 &   7.59 \\
         & HWDM &  2.10 &   7.58 \\
\bottomrule
\end{tabular}
\end{table}

\subsection{Binary performance}
Binary performance results as measured by AUC are provided in Table \ref{tab:binary_performance}. In addition, unadjusted and more detailed metrics are provided in the Appendix \ref{unadjusted_pffh} and \ref{unadjusted_pffo}. The AUCs of one-step ahead PFFH predictions were high (0.98-0.99) for all the forecasting models. The accuracy of the ETS models decreased monotonically as a function of forecast horizon and bottomed at 0.79 at $t+24$. PFFO was low at 0 a.m. with AUC ranging between 0.47-0.50 but the performance increased gradually over the course of the day. HWDM reached an AUC 0.70 at 11 a.m., 0.78 at 12 a.m., 0.84 at 1 p.m. and finally 0.87 at 3 p.m. Differences between HWDM and other ETS models were small.

\begin{table}[h]
\centering
\caption{The performance of the models over different forecast horizons (PFFH) and forecast origins (PFFO) reported with Area Under Curve (AUC).}
\label{tab:binary_performance}
\begin{tabular}{rcccrccc}
\toprule
   PFFH & \multicolumn{4}{r}{PFFO} \\
Horizon & AHWM & HWDM & MHWM & Origin & AHWM & HWDM & MHWM \\
\midrule
    t+1 & 0.99 & 0.98 & 0.99 &      0 & 0.50 & 0.52 & 0.50 \\
    t+2 & 0.97 & 0.97 & 0.97 &      1 & 0.48 & 0.50 & 0.50 \\
    t+3 & 0.96 & 0.96 & 0.96 &      2 & 0.49 & 0.54 & 0.52 \\
    t+4 & 0.95 & 0.94 & 0.95 &      3 & 0.49 & 0.54 & 0.55 \\
    t+5 & 0.94 & 0.93 & 0.93 &      4 & 0.50 & 0.54 & 0.55 \\
    t+6 & 0.93 & 0.91 & 0.92 &      5 & 0.53 & 0.57 & 0.54 \\
    t+7 & 0.92 & 0.90 & 0.91 &      6 & 0.53 & 0.58 & 0.58 \\
    t+8 & 0.91 & 0.89 & 0.90 &      7 & 0.56 & 0.59 & 0.61 \\
    t+9 & 0.91 & 0.89 & 0.89 &      8 & 0.65 & 0.64 & 0.66 \\
   t+10 & 0.91 & 0.89 & 0.89 &      9 & 0.65 & 0.68 & 0.64 \\
   t+11 & 0.90 & 0.87 & 0.88 &     10 & 0.61 & 0.64 & 0.64 \\
   t+12 & 0.87 & 0.86 & 0.86 &     11 & 0.68 & 0.73 & 0.72 \\
   t+13 & 0.85 & 0.85 & 0.85 &     12 & 0.76 & 0.79 & 0.79 \\
   t+14 & 0.84 & 0.83 & 0.83 &     13 & 0.80 & 0.84 & 0.83 \\
   t+15 & 0.82 & 0.81 & 0.81 &     14 & 0.85 & 0.85 & 0.86 \\
   t+16 & 0.81 & 0.80 & 0.79 &     15 & 0.89 & 0.88 & 0.89 \\
   t+17 & 0.81 & 0.79 & 0.80 &     16 & 0.85 & 0.84 & 0.84 \\
   t+18 & 0.81 & 0.79 & 0.79 &     17 & 0.79 & 0.79 & 0.77 \\
   t+19 & 0.81 & 0.79 & 0.79 &     18 & 0.76 & 0.74 & 0.75 \\
   t+20 & 0.82 & 0.80 & 0.81 &     19 & 0.62 & 0.60 & 0.60 \\
   t+21 & 0.82 & 0.80 & 0.81 &     20 & 0.59 & 0.57 & 0.59 \\
   t+22 & 0.81 & 0.80 & 0.81 &     21 & 0.63 & 0.61 & 0.61 \\
   t+23 & 0.80 & 0.79 & 0.80 &     22 & 0.56 & 0.56 & 0.55 \\
   t+24 & 0.79 & 0.79 & 0.79 &     23 & 0.55 & 0.54 & 0.54 \\
\bottomrule
\end{tabular}
\end{table}

While AUC is calculated over all thresholds, there are metrics, such as F1, that are calculated with a single threshold. While the models seem to have equal accuracy in terms of AUC, F1 shows clear differences: In comparison to HWDM and MHWM, AHWM performs relatively poorly when the forecast horizon is longer than four hours or when the forecast origin is between 0 a.m. and 3 p.m. (see Figures \ref{fig:horizonwise_metrics} and \ref{fig:originwise_metrics} in Appendix). This is because in these cases, the sensitivity (also known as recall and true positive rate) is low, meaning that only a small fraction of crowding cases were correctly predicted in these cases.

\section{Discussion}\label{discussion}
In this study, we had three main findings. First, we showed that it is possible to build a prospective ED crowding early warning software using simple statistical forecasting models and with very limited resources. Second, we showed that even simple univariate models can provide excellent binary accuracy and potentially provide actionable information to ED stakeholders with modest computing and software requirements. Third, binary performance is significantly affected by forecast origin which should be considered when performance metrics are reported.

PFFH was high regardless of the forecast horizon with AUC ranging between 0.77-0.98. There is a clear discrepancy with these accuracies compared to PFFO results with AUC ranging from 0.47-0.85. This difference becomes understandable when it is weighed against the ideas introduced in Section \ref{introduction}. In addition, PFFH results are averages over all the predictions made at different forecast origins. This results in overly optimistic picture of the model performance. For example, consider predictions made between 0 a.m. and 13 a.m. during of which the ED was not crowded even once (Figure \ref{fig:crowding_dist_hourly}). During these hours, even a simple seasonal naive model would provide perfect $t+24$ classification accuracy (and identically perfect $t+h$ accuracy as long as $t+h < 13$). This effectively "inflates" the resulting performance, which is especially misleading because the administration is not particularly interested in the occupancy levels of these specific hours, which are known to be relatively quiet.

For these reasons, the PFFO results provide a better and more truthful insight into the actual potential of the models. As shown in Table \ref{tab:binary_performance}, the models had no discrimatory power at 0 a.m. but the AUC gradually increased over the course of the day, reaching acceptable level of 0.70 at 11 a.m. and excellent level of 0.84 at 1 p.m. This improvement was expected because, in this setting, the effective forecast horizon decreases as a function of increasing forecast origin. This means that the models are iteratively presented with the most recent occupancy statistics and are able to correct the prediction based on the status of the ED at prediction time. In fact, looking at the unadjusted accuracy metrics in Figure \ref{fig:originwise_metrics}, the specificity of the ETS models increases over the course of the day whereas sensitivity remains constant. Regardless, the system reached an acceptable level of accuracy as early as at 11 a.m. If the predictions were used to guide administrative decisions and since the vast majority of crowding events were observed after 6 p.m. as shown in Figure \ref{fig:crowding_dist_hourly}, there would have been a several hour margin of action to perform any number of preventive measures.

Further, it would be interesting to evaluate the performance of these models when trained on daily resolution. For example, in terms of forecasting crowding, the real targets of interest are not the hourly occupancy statistics, but the peak occupancy of the afternoon. The same models could potentially perform better, if trained with series of these daily peak occupancies, which likely demonstrate clear weekly seasonality and potential trends. Moreover, it would be likely beneficial to estimate classification techniques, such as logistic regression or support vector machines. After all, we are only marginally interested in the underlying numbers of the patients and very much interested in the binary future state of the ED. We presume that the use of classification techniques has the potential to enhance significantly the sensitivity of the system.

The focus of this study was deliberately on building the software and integrating it into hospital information infrastructure, which was not a trivial task. For this reason, the models used were simple well-established statistical models. However, several novel machine learning based models have been introduced over the course of the last few years, and these models would likely provide better accuracy in this context.

\section{Conclusions}\label{conclusions}
To conclude, we showed the ability of a prospective overcrowding early warning system to predict next hour overcrowding with a nominal AUC of 0.98 and 24 hour overcrowding with an AUC of 0.79. We also suggest that reporting accuracies in this manner is highly misleading and propose a clinically oriented PFFO metric. Using this approach, we suggest that afternoon overcrowding can be predicted at 1 p.m. with an AUC of 0.84.

Future work should 1) investigate the use of binary cost function optimally with a machine learning model, 2) investigate the accuracy of the system with stratified sample, 3) perform a pilot with the ED staff to validate the operational benefits of the system in clinical setting and 4) document the binary performance of other previously proposed ED forecasting models.

\section*{Acknowledgements}
This study was funded by the Doctoral Programme in Medicine and Life Sciences, Faculty of Medicine and Health Technology, Tampere University; Finnish Society of Emergency Medicine; Hauho Oma Savings Bank Foundation and Renko Oma Savings Bank Foundation. We acknowledge Pirkanmaa Hospital District (Emil Ackerman) for financing the prototype and Istekki Oy (Mikko Kuru) for establishing and maintaining the virtual machine.

\bibliographystyle{unsrt}
\bibliography{references}

\begin{thebibliography}{10}

\bibitem{Richardson2006}
Drew~B. Richardson.
\newblock {Increase in patient mortality at 10 days associated with emergency
  department overcrowding}.
\newblock {\em Medical Journal of Australia}, 184(5):213--216, 2006.

\bibitem{Berg2019}
Lena~M. Berg, Anna Ehrenberg, Jan Florin, Jan {\"{O}}stergren, Andrea
  Discacciati, and Katarina~E. G{\"{o}}ransson.
\newblock {Associations Between Crowding and Ten-Day Mortality Among Patients
  Allocated Lower Triage Acuity Levels Without Need of Acute Hospital Care on
  Departure From the Emergency Department}.
\newblock {\em Annals of Emergency Medicine}, 74(3):345--356, 2019.

\bibitem{Jo2014}
Sion Jo, Young~Ho Jin, Jae~Baek Lee, Taeoh Jeong, Jaechol Yoon, and Boyoung
  Park.
\newblock {Emergency department occupancy ratio is associated with increased
  early mortality}.
\newblock {\em Journal of Emergency Medicine}, 46(2):241--249, 2014.

\bibitem{Guttmann2011}
Astrid Guttmann, Michael~J. Schull, Marian~J. Vermeulen, and Therese~A. Stukel.
\newblock {Association between waiting times and short term mortality and
  hospital admission after departure from emergency department: Population
  based cohort study from Ontario, Canada}.
\newblock {\em Bmj}, 342(7809), 2011.

\bibitem{Sprivulis2006}
Frazer~A {Sprivulis P.C, Da Silva J.A., Jacobs I.} and Jelinek G.
\newblock {The association between hospital overcrowding and mortality among
  patients admitted via Western Australian emergency departments}.
\newblock {\em Mja}, 184(5):208--212, 2006.

\bibitem{Janke2022}
Alexander~T. Janke, Edward~R. Melnick, and Arjun~K. Venkatesh.
\newblock {Hospital Occupancy and Emergency Department Boarding During the
  COVID-19 Pandemic}.
\newblock {\em JAMA Network Open}, 5(9):e2233964--e2233964, 09 2022.

\bibitem{Asplin2003}
Brent~R. Asplin, David~J. Magid, Karin~V. Rhodes, Leif~I. Solberg, Nicole
  Lurie, and Carlos~A. Camargo.
\newblock {A conceptual model of emergency department crowding}.
\newblock {\em Annals of Emergency Medicine}, 42(2):173--180, 2003.

\bibitem{Morley2018}
Claire Morley, Maria Unwin, Gregory~M. Peterson, Jim Stankovich, and Leigh
  Kinsman.
\newblock {Emergency department crowding: A systematic review of causes,
  consequences and solutions}.
\newblock 13(8):1--42, 2018.

\bibitem{Gul2020}
Muhammet Gul and Erkan Celik.
\newblock An exhaustive review and analysis on applications of statistical
  forecasting in hospital emergency departments.
\newblock {\em Health Systems}, 9(4):263--284, 2020.

\bibitem{Caldas2022}
Francisco~M. Caldas and Cláudia Soares.
\newblock A temporal fusion transformer for long-term explainable prediction of
  emergency department overcrowding, 2022.

\bibitem{Harrou2020}
Fouzi Harrou, Abdelkader Dairi, Farid Kadri, and Ying Sun.
\newblock {Forecasting emergency department overcrowding: A deep learning
  framework}.
\newblock {\em Chaos, Solitons and Fractals}, 139:110247, 2020.

\bibitem{Sharafat2021}
Ali~R. Sharafat and Mohsen Bayati.
\newblock {PatientFlowNet: A Deep Learning Approach to Patient Flow Prediction
  in Emergency Departments}.
\newblock {\em IEEE Access}, 9:45552--45561, 2021.

\bibitem{Fan2022}
Bi~Fan, Jiaxuan Peng, Hainan Guo, Haobin Gu, Kangkang Xu, and Tingting Wu.
\newblock Accurate forecasting of emergency department arrivals with internet
  search index and machine learning models: Model development and performance
  evaluation.
\newblock {\em JMIR Med Inform}, 10(7):e34504, Jul 2022.

\bibitem{Elstein2002}
Arthur~S Elstein and Alan Schwarz.
\newblock Clinical problem solving and diagnostic decision making: selective
  review of the cognitive literature.
\newblock {\em BMJ}, 324(7339):729--732, 2002.

\bibitem{Ekstrom2015}
Andreas Ekstr{\"{o}}m, M~Ed, Lisa Kurland, Nasim Farrokhnia, Maaret
  Castr{\'{e}}n, and Martin Nordberg.
\newblock {Forecasting Emergency Department Visits Using Internet Data}.
\newblock {\em Annals of Emergency Medicine}, 65(4):436--442.e1, 2015.

\bibitem{Asheim2019}
Andreas Asheim, Lars~P. {Bache-Wiig Bj{\o}rnsen}, Lars~E. N{\ae}ss-Pleym,
  Oddvar Uleberg, Jostein Dale, and Sara~M. Nilsen.
\newblock {Real-time forecasting of emergency department arrivals using
  prehospital data}.
\newblock {\em BMC Emergency Medicine}, 19(1):1--6, 2019.

\bibitem{Winters1960}
Peter~R Winters.
\newblock Forecasting sales by exponentially weighted moving averages.
\newblock {\em Management Science}, 6:324--342, 1960.

\bibitem{statsmodels}
Skipper Seabold and Josef Perktold.
\newblock statsmodels: Econometric and statistical modeling with python.
\newblock In {\em 9th Python in Science Conference}, 2010.

\bibitem{forman2010apples}
George Forman and Martin Scholz.
\newblock Apples-to-apples in cross-validation studies: pitfalls in classifier
  performance measurement.
\newblock {\em Acm Sigkdd Explorations Newsletter}, 12(1):49--57, 2010.

\end{thebibliography}

\clearpage
\appendix

\section{Appendix}

\subsection{Missing data}\label{missing_data}
\begin{figure}[H]
     \centering
     \begin{subfigure}[b]{1.00\textwidth}
         \centering
         \includegraphics[width=\textwidth]{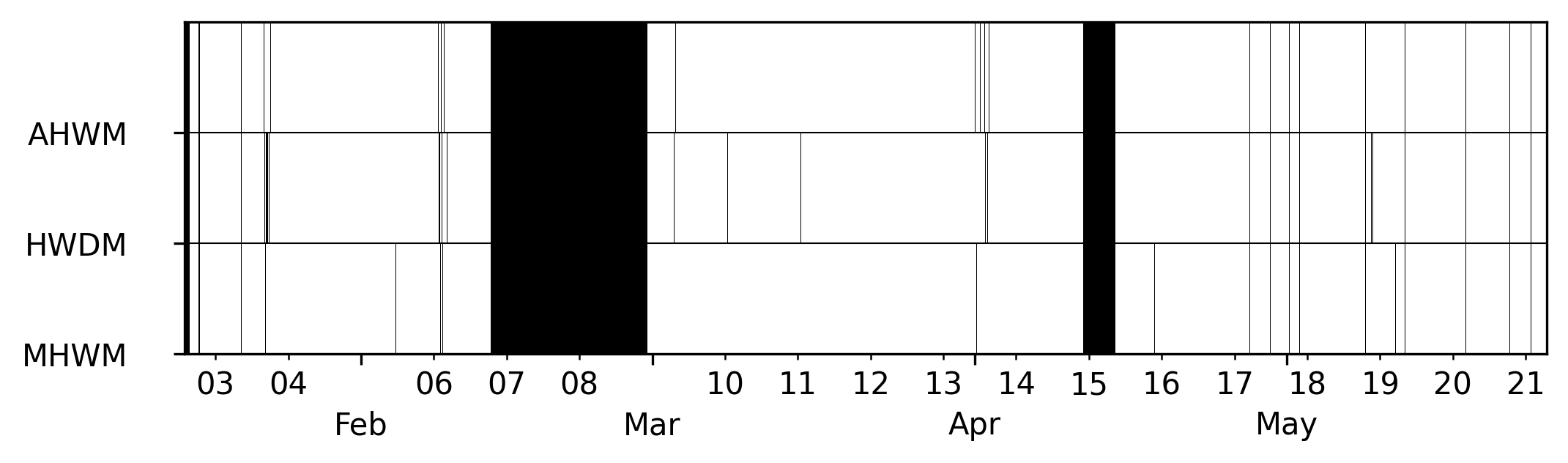}
         \caption{Arrivals}
         \label{fig:missing_data_hourly_arrivals}
     \end{subfigure}
     \hfill
     \begin{subfigure}[b]{1.00\textwidth}
         \centering
         \includegraphics[width=\textwidth]{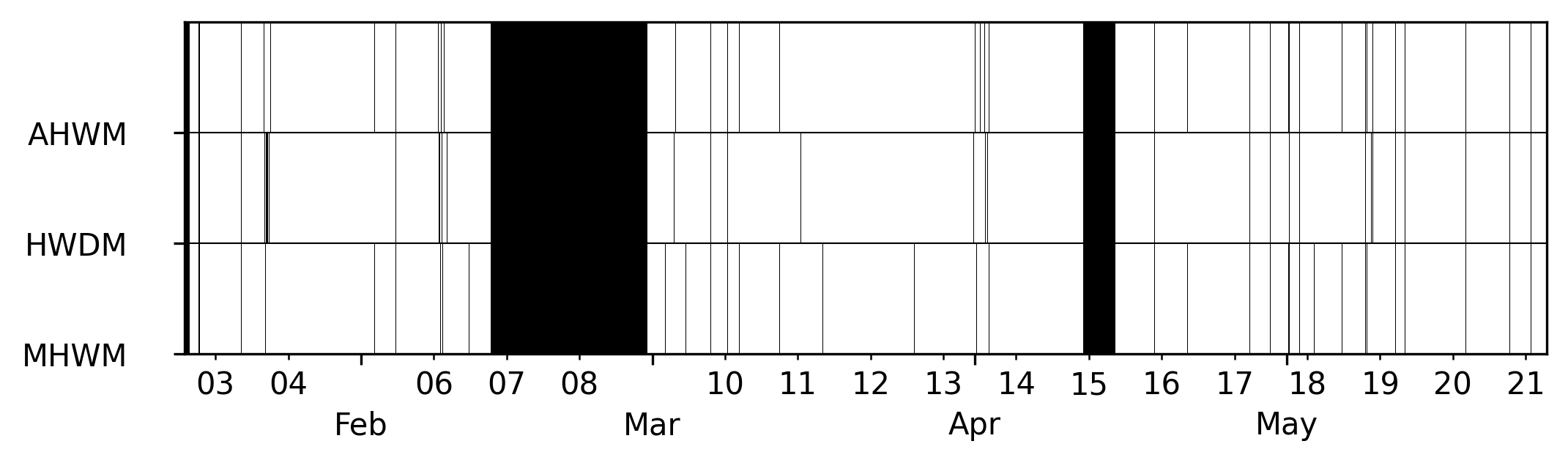}
         \caption{Occupancy}
         \label{fig:Occupancy}
     \end{subfigure}
     \caption{Temporal distribution of missing data, marked with black color. Two week period of missing data during weeks seven and eight are due to a credentials issue.}
     \label{fig:missing_data}
\end{figure}

\clearpage

\subsection{Unadjusted PFFH metrics}\label{unadjusted_pffh}

\begin{figure}[h]
    \centering
    \begin{subfigure}[b]{0.49\textwidth}
        \centering
        \includegraphics[width=\textwidth]{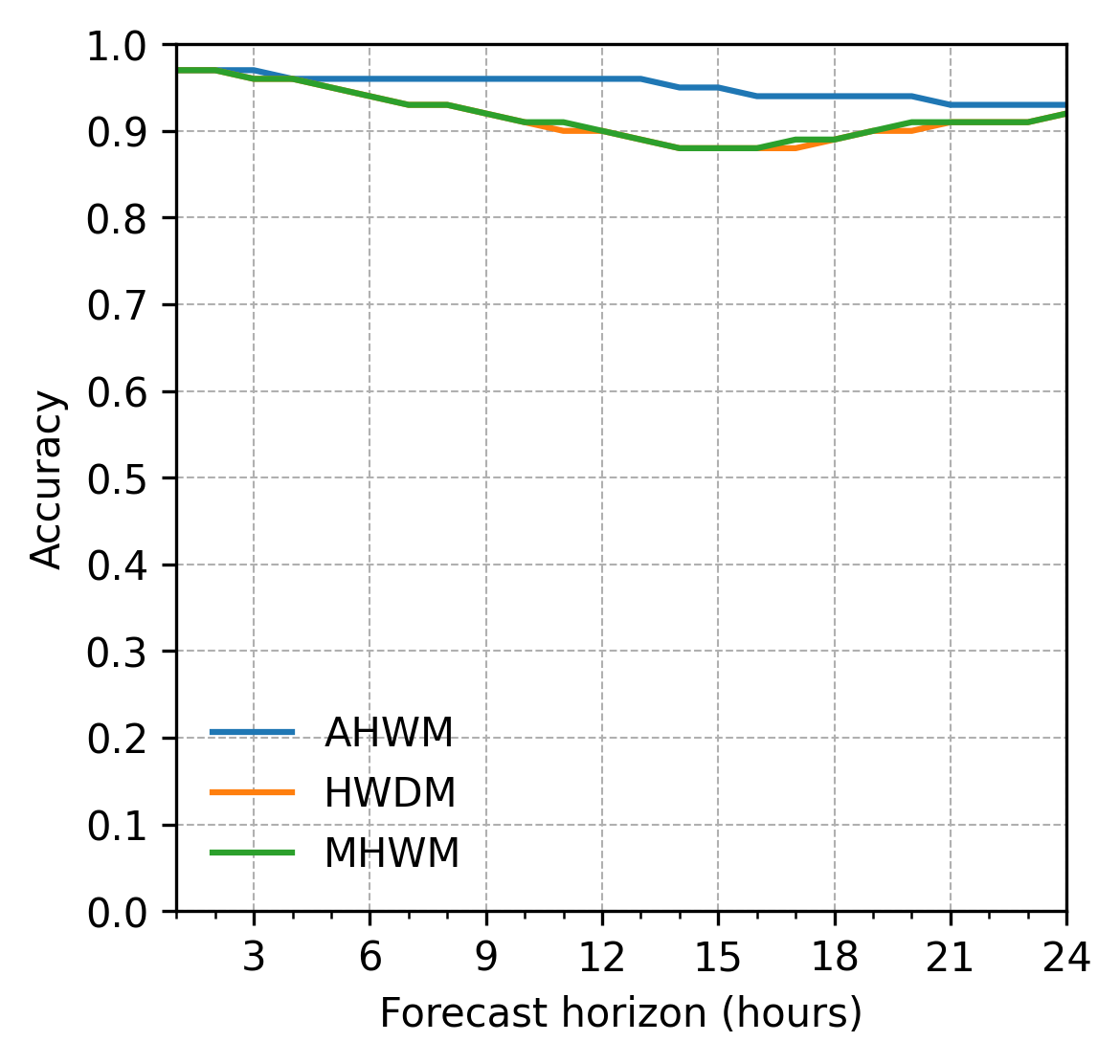}
        \caption{Accuracy}
        \label{fig:horizonwise_accuracy}
    \end{subfigure}
    \hfill
    \begin{subfigure}[b]{0.49\textwidth}
        \centering
        \includegraphics[width=\textwidth]{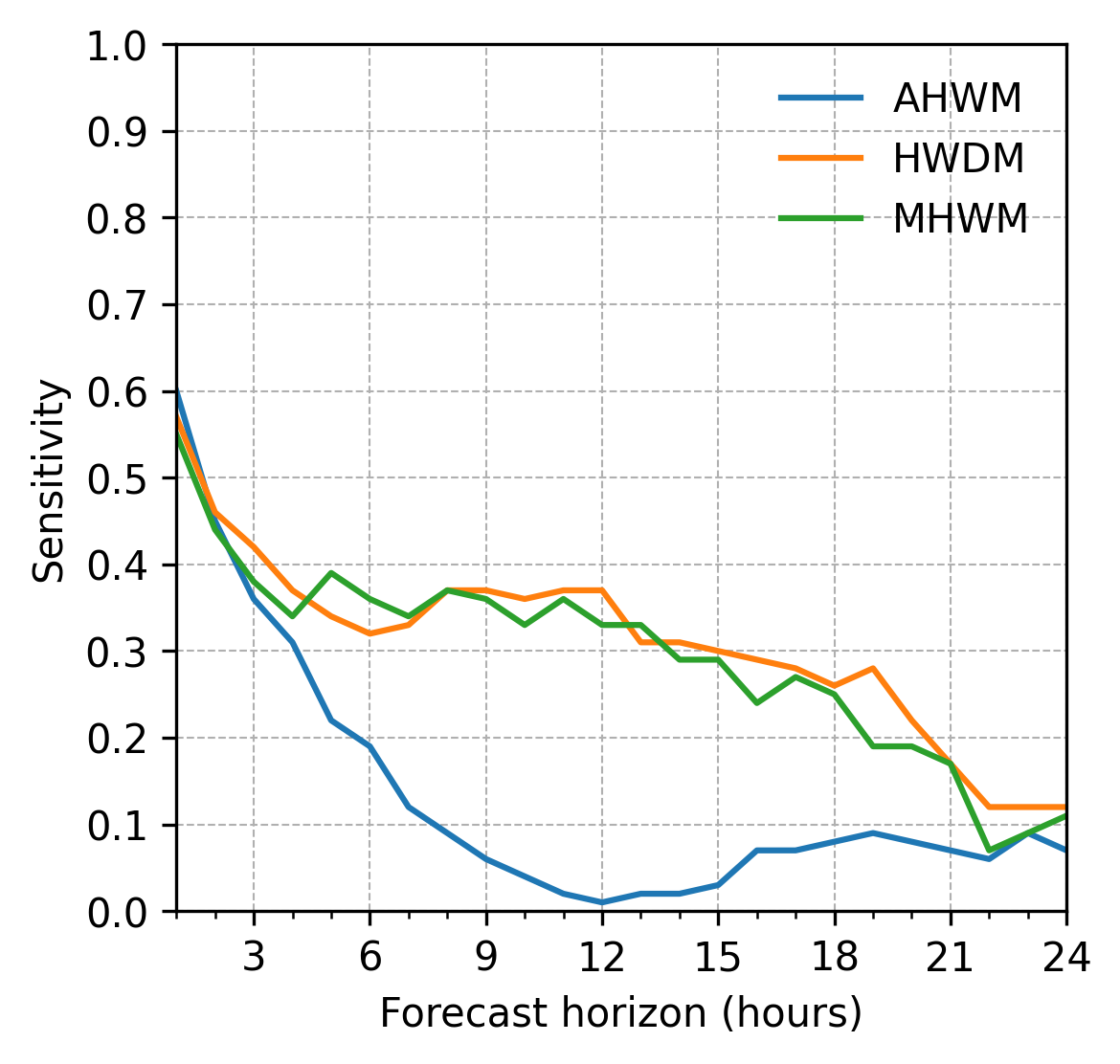}
        \caption{Sensitivity}
        \label{fig:horizonwise_sensitivity}
    \end{subfigure}
    \hfill
    \begin{subfigure}[b]{0.49\textwidth}
        \centering
        \includegraphics[width=\textwidth]{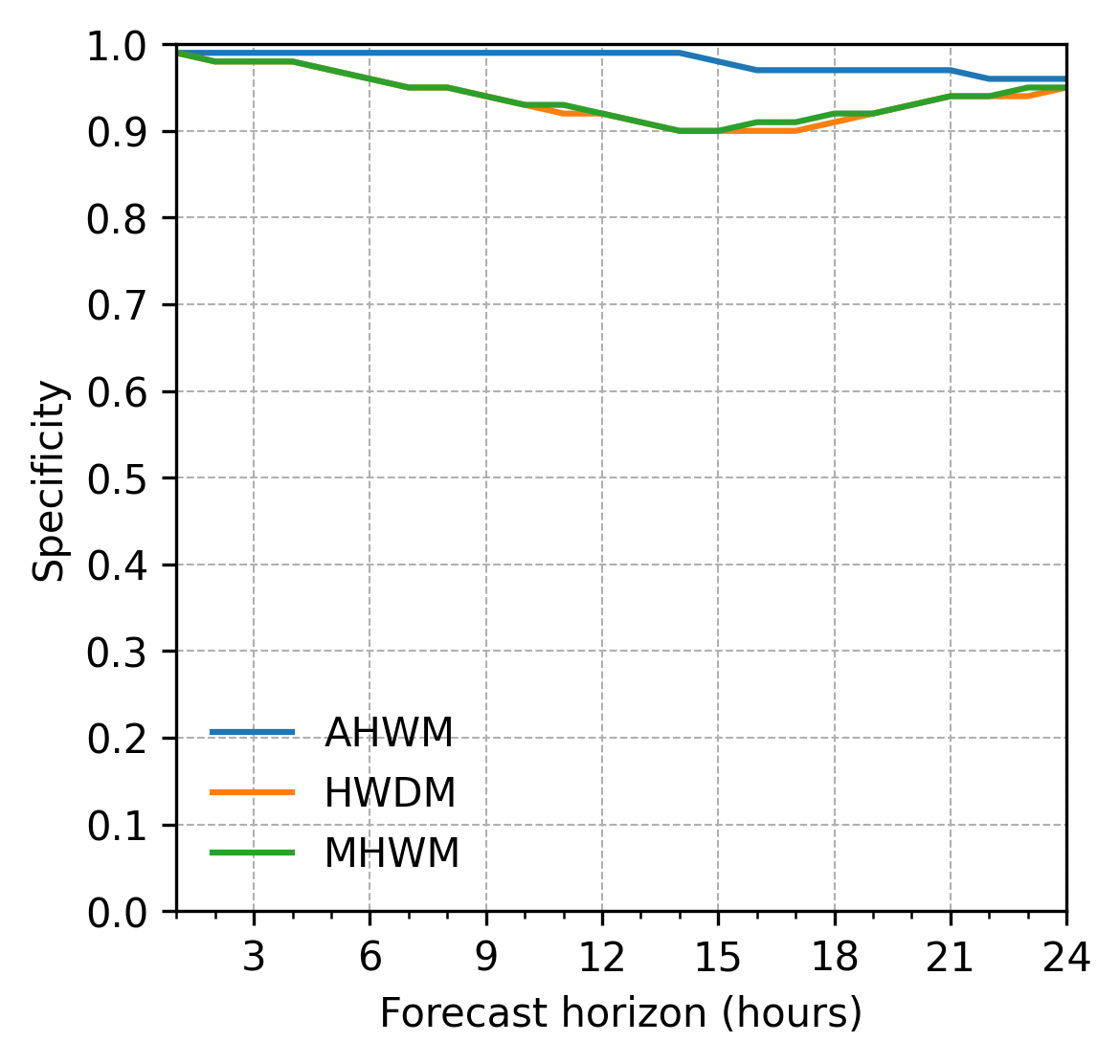}
        \caption{Specificity}
        \label{fig:horizonwise_specificity}
    \end{subfigure}
    \hfill
    \begin{subfigure}[b]{0.49\textwidth}
        \centering
        \includegraphics[width=\textwidth]{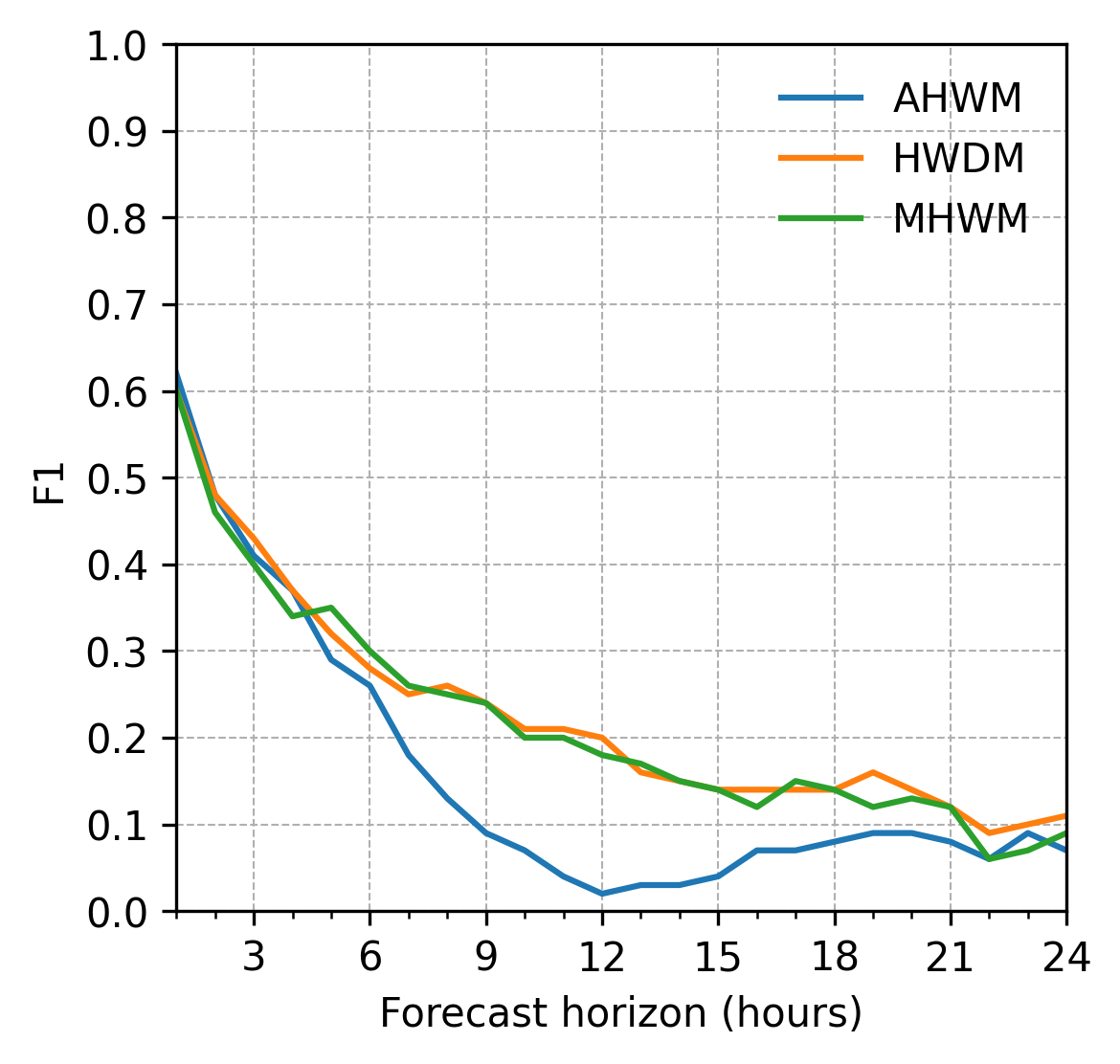}
        \caption{F1}
        \label{fig:horizonwise_f1}
    \end{subfigure}
        \caption{Performance as function of forecast horizon}
        \label{fig:horizonwise_metrics}
\end{figure}

Unadjusted performance metrics as a function of forecast horizon are shown in Figure \ref{fig:horizonwise_metrics}. Overall accuracy of the models was high and nearly identical ranging between 88-97\%. This was mostly explained by equally high specificity of 90-100\%. AHWM had the highest one-step ahead sensitivity of 60\%, but the difference compared to other models was small. The sensitivity declined as a function of increasing forecast horizon, but this effect was more prononounced with HWDM compared to AHWM and HWDM.

\subsection{Unadjusted PFFO metrics}\label{unadjusted_pffo}
\begin{figure}[h]
     \centering
     \begin{subfigure}[b]{0.49\textwidth}
         \centering
         \includegraphics[width=\textwidth]{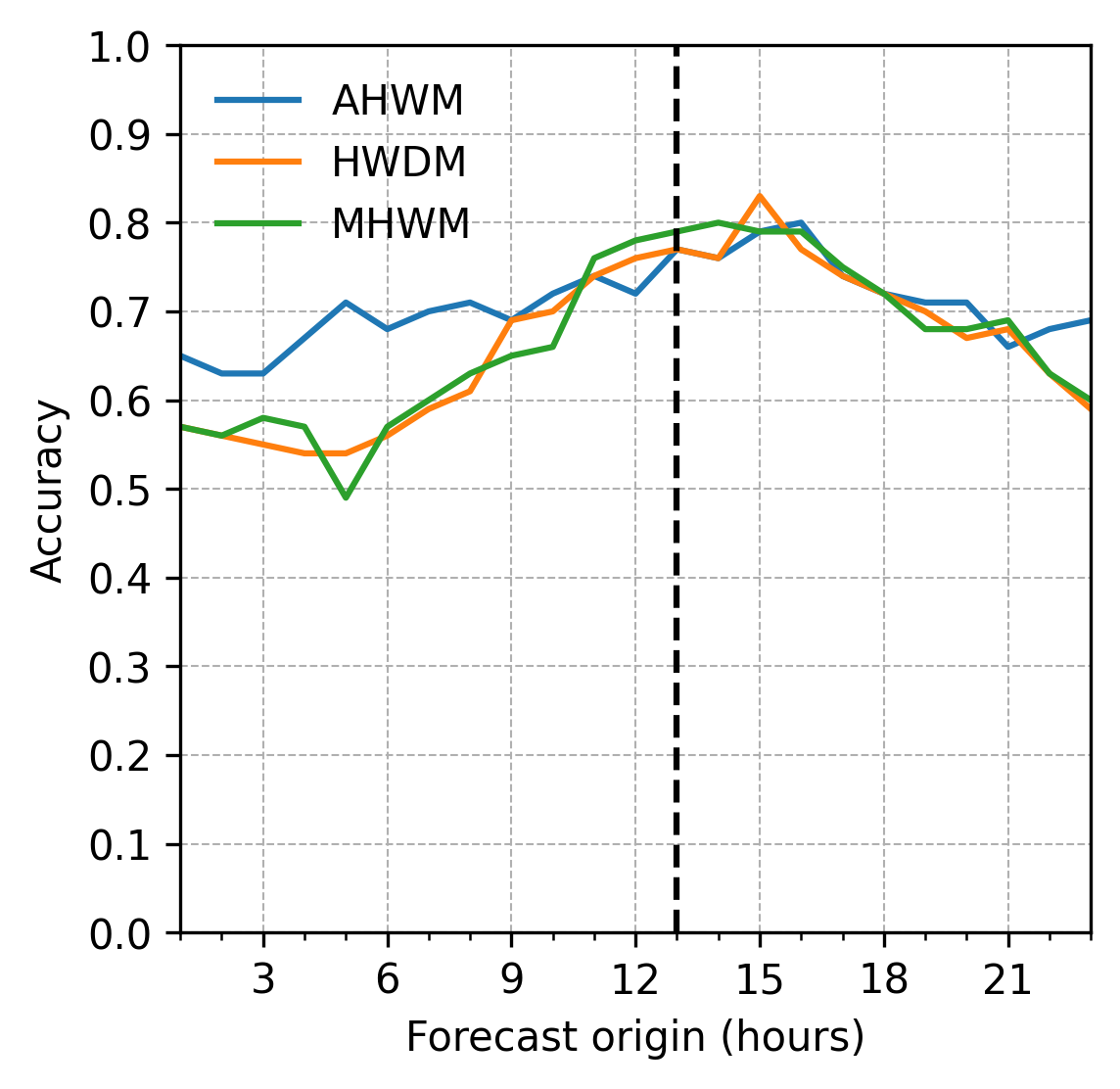}
         \caption{Accuracy}
         \label{fig:originwise_accuracy}
     \end{subfigure}
     \hfill
     \begin{subfigure}[b]{0.49\textwidth}
         \centering
         \includegraphics[width=\textwidth]{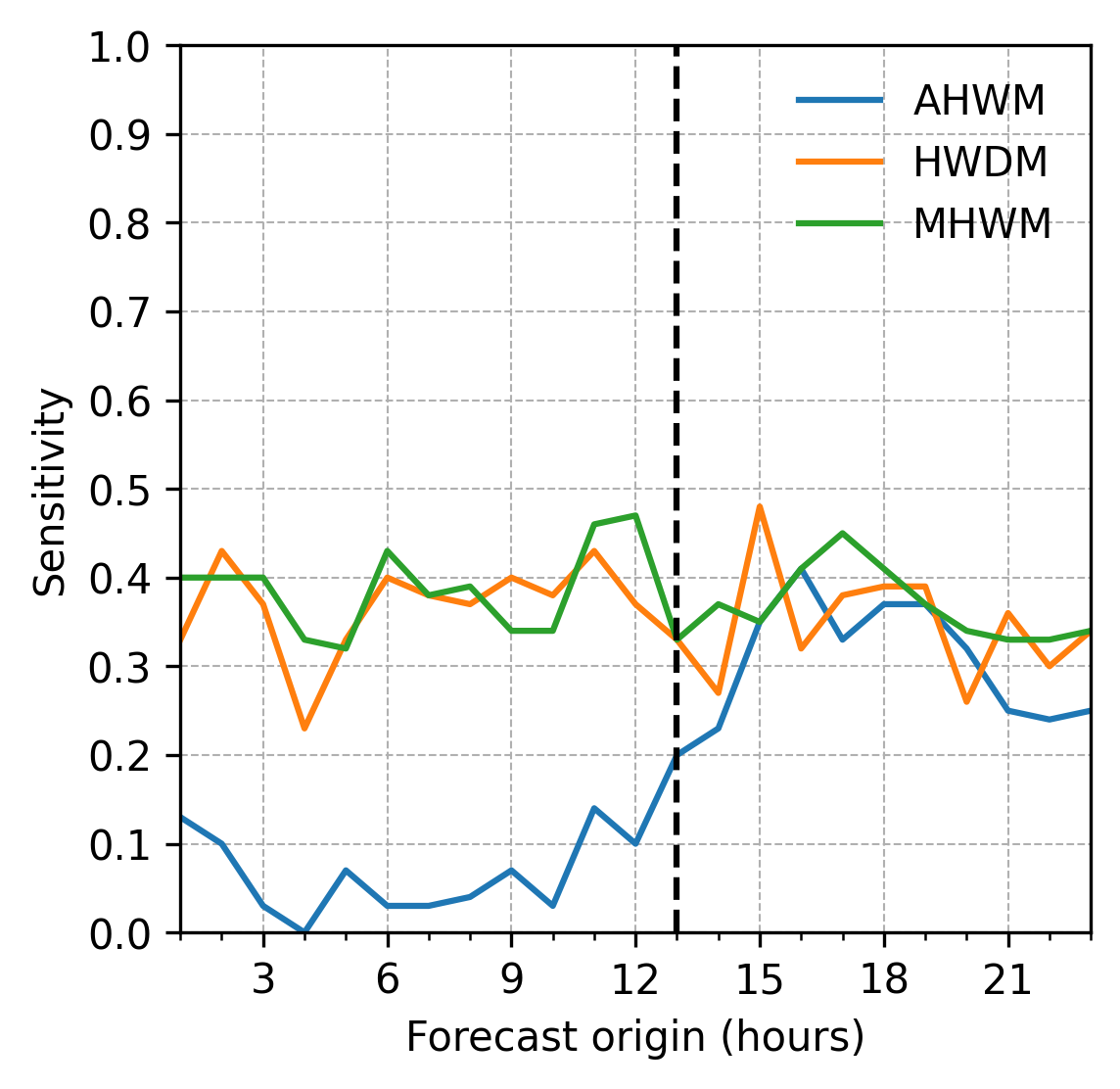}
         \caption{Sensitivity}
         \label{fig:originwise_sensitivity}
     \end{subfigure}
     \hfill
     \begin{subfigure}[b]{0.49\textwidth}
         \centering
         \includegraphics[width=\textwidth]{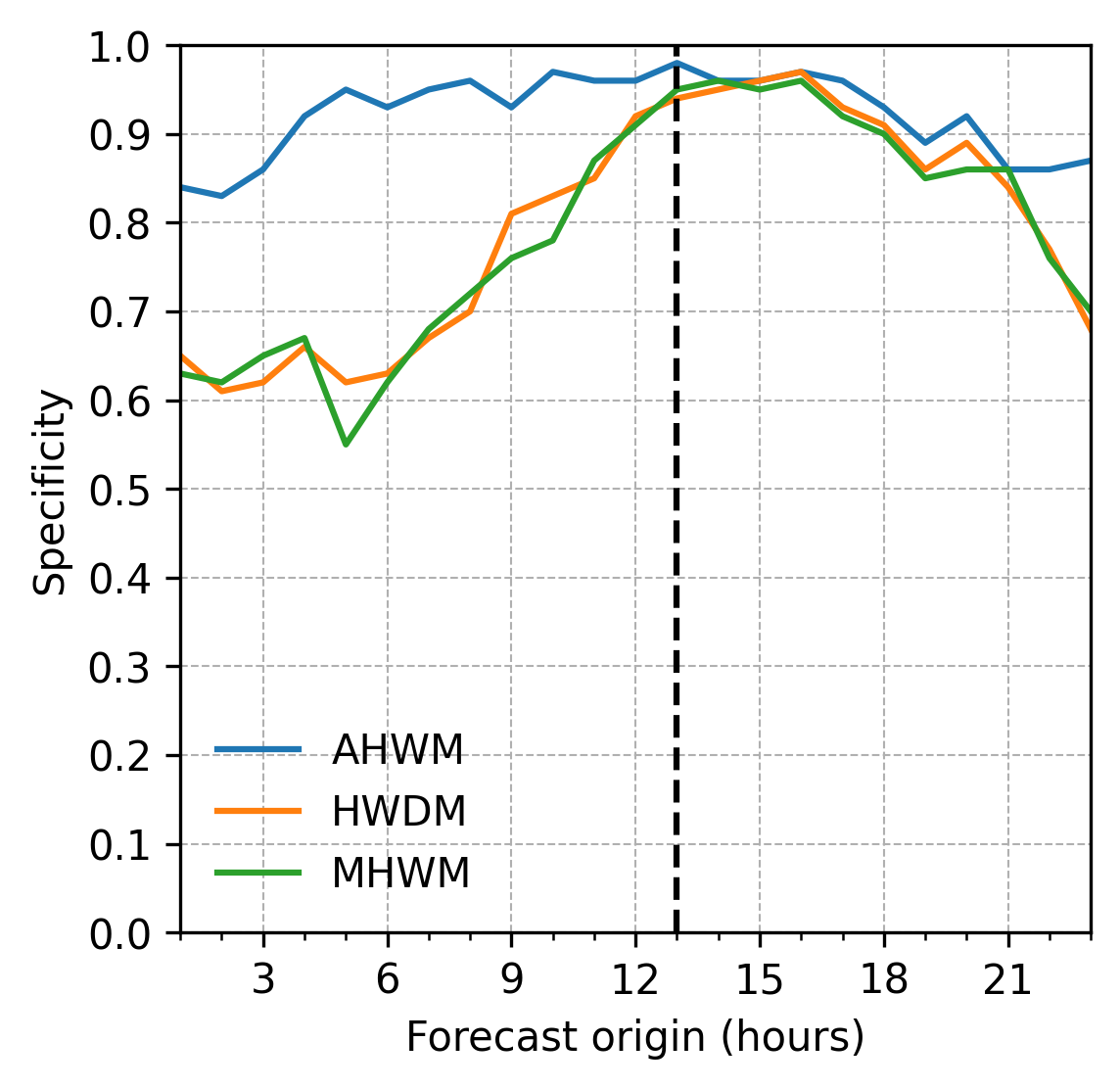}
         \caption{Specificity}
         \label{fig:originwise_specificity}
    \end{subfigure}
     \hfill
     \begin{subfigure}[b]{0.49\textwidth}
         \centering
         \includegraphics[width=\textwidth]{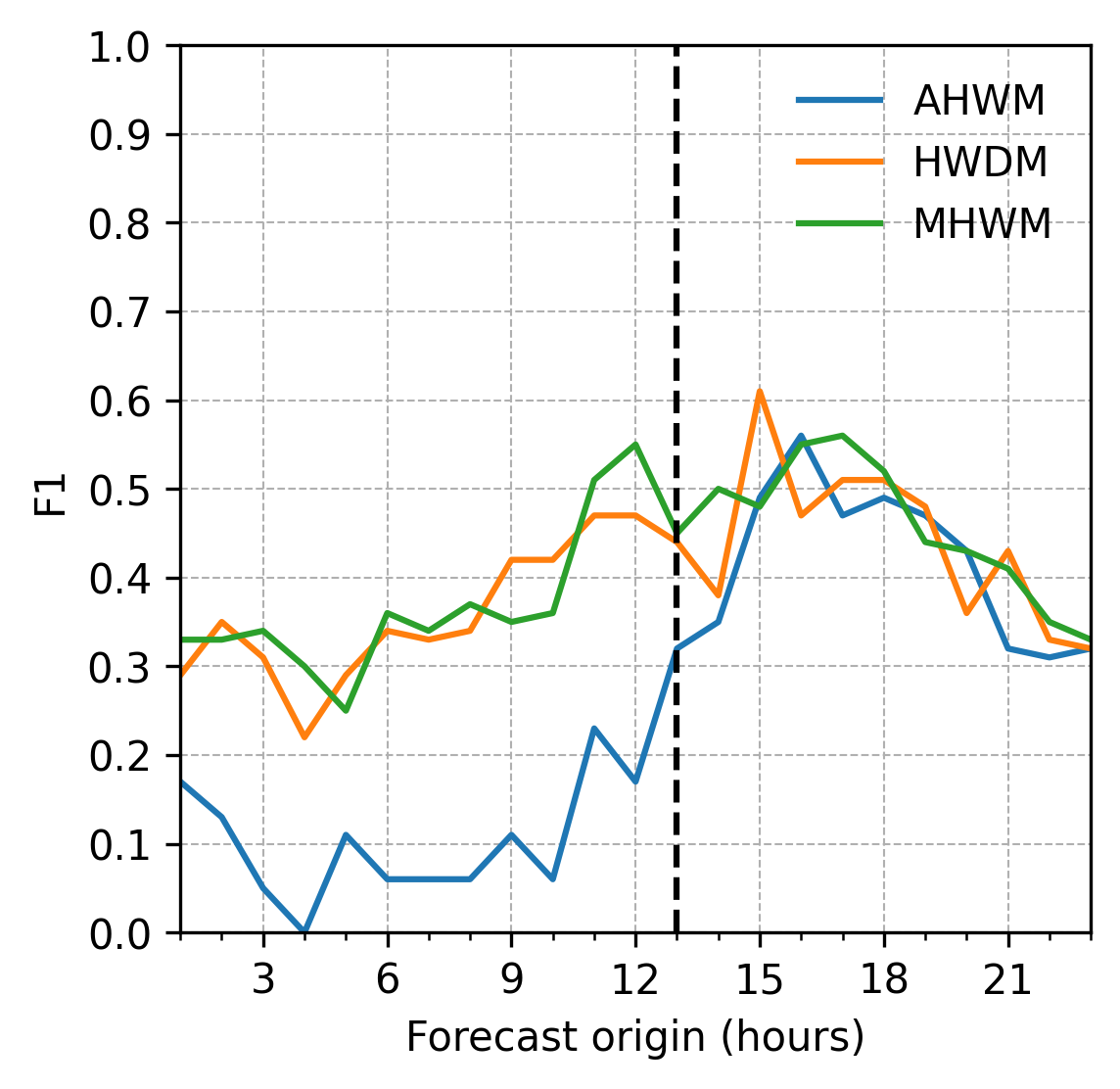}
         \caption{F1}
         \label{fig:originwise_f1}
     \end{subfigure}
        \caption{Performance as function of forecast origin}
        \label{fig:originwise_metrics}
\end{figure}

Unadjusted performance metrics as a function of forecast origin are shown in Figure \ref{fig:originwise_metrics}. The sensitivity of AHWM was low, ranging between 0\% and 20\% before 1 p.m. HWDM and MHWM demonstrated identical and relatively constant sensitivity irrespective of the forecast origin with mean sensitivity of 35-38\%. The specificity of AHWM was high and relatively constant with values ranging with means of 98\% and 92\% respectively. The specificity of HWDM and MHWM increased as a function of forecast origin from 63-65\% at 1 a.m. to 94-95\% at 1 p.m.

\end{document}